\documentstyle[12pt,aps]{revtex}

\tightenlines

\draft

\begin{document}

\large

\title{On the energy translation invariance of probability distributions}

\author{Q.A. Wang$^1$\footnote{Corresponding author : awang@ismans.univ-lemans.fr},
L. Nivanen$^1$, M. Pezeril$^2$ and A. Le M\'ehaut\'e$^1$}
\address{$^1$Institut Sup\'{e}rieur des Mat\'{e}riaux du Mans, 44, Av.
Bartholdi, 72000 Le Mans, France}

\address{$^2$Laboratoire de Physique de l'\'Etat Condens\'e, Universit\'e du Maine,
72000 Le Mans, France }

\maketitle

\begin{abstract}
We comment on the problem of energy translation invariance of probability
distribution and present some observations. It is shown that a probability
distribution can be invariant in the thermodynamic limit if there is no long
term interaction or correlation and no relativistic effect. So this invariance
should not be considered as a universal theoretical property. Some
peculiarities within the invariant $q$-exponential distribution reveal that the
connection of the current nonextensive statistical mechanics to thermodynamics
might be disturbed by this invariance.
\end{abstract}

\pacs{02.50.-r,05.20.-y,05.30.-d,05.70.-a}

\section{Introduction}
The nonextensive statistical mechanics
(NSM)\cite{Tsal1,Tsal2,Tsal3,Penni,Wang00,Wang01,Wang02} is a generalization of
the Boltzmann-Gibbs statistics (BGS) intended to study complex nonlinear
systems having long-range correlations and fractal or chaotic state space for
which BGS is no more valid. Due to the generalized character of this theory,
the scientists of this field have been brought back to reflections about many
fundamental aspects of physics theory. Can we keep the additivity of, e.g.,
entropy, energy and volume for complex systems? What are the possible relation
between the probabilities of interacting systems? What is thermodynamic
equilibrium in this case? How to define it for systems of same nonextensive
nature and for systems of different nonextensive nature? Is the classical
probability definition still valid in chaotic or fractal phase space? Does the
completeness of information about systems containing independent parts still
apply to systems with correlated or overlapped parts ... ? All these questions
concerning the foundation of NSM are far from being answered and it would be
too soon to conclude. A complete comprehension of these aspects is necessary to
finally determine the validity limites of NSM and to know the kind of systems
to which NSM should be applied.

One of the most worrying problems of the NSM distribution given by
\begin{equation} \label{a1}
p_i=\frac{1}{Z}[1-(1-q)\beta e_i]^\frac{1}{1-q}
\end{equation}
is that it is not invariant under uniform translation of the energy spectra
$e_i$\cite{Tsal3,Sist99}. Only when $q=1$, Eq.(\ref{a1}) becomes
$p_i=\frac{1}{Z}e^{-\beta e_i}$ and the invariance can be recovered. This
$q$-exponential distribution can be given by maximizing Tsallis
entropy\footnote{Tsallis entropy is given by
$S=-k\frac{\sum_{i=1}^{w}p_i-\sum_{i=1}^{w}p_i^q}{1-q} , (q \in
R)$\cite{Tsal1}} with the unnormalized expectation\cite{Tsal2}
\begin{equation} \label{aa1}
U=\sum_{i}^we_ip_i^q.
\end{equation}
and the normalization $\sum_{i}^wp_i=1$ from which the partition function $Z$
is calculated. Eq.(\ref{a1}) can also be found, with a different partition
function, by maximizing Tsallis entropy with Eq.(\ref{aa1}) and an anomalous
normalization $\sum_{i}^wp_i^q=1$\cite{Wang00}. $\beta$ can be identified to
the inverse temperature within a nonadditive energy
scenario\cite{Wang00,Wang01,Wang02,Wang02a} or with an additive energy
approximation\cite{Abe99,Mart01}).

From the power law distribution Eq.(\ref{a1}), it is obvious that,
if we replace $e_i$ by $e_i+C$ where $C$ is constant, we find
$p_i(e_i+C)\neq p_i(e_i)$, excepted that $q=1$. If $p_i$ varies
with $C$, then the variance of all the thermodynamic functions
will be disturbed by $C$. For example, for the energy U, we have
\begin{eqnarray} \label{aa4}
U(e+C) &=& \sum_i(e_i+C)p_i^q(e_i+C) \\ \nonumber
&=&\frac{1}{Z^q(e+C)}\sum_i(e_i+C)[1-(1-q)\beta
(e_i+C)]^\frac{q}{1-q}\\ \nonumber
&=&\frac{-1}{Z^q(e+C)}\frac{\partial}{\partial\beta}\sum_i[1-(1-q)\beta
(e_i+C)]^\frac{1}{1-q}\\ \nonumber
&=&\frac{-1}{Z^q(e+C)}\frac{\partial Z(e+C)}{\partial\beta}\\
\nonumber
&=&-\frac{\partial}{\partial\beta}\frac{Z^{1-q}(e+C)-1}{1-q}\\
\nonumber
&\neq&U+C
\end{eqnarray}
where $U=-\frac{\partial}{\partial\beta}\frac{Z^{1-q}-1}{1-q}$ is the
unnormalized expectation without energy translation ($C=0$). This theoretical
feature seems difficult to accept according to some
scientists\cite{Tsal3,Sist99}. Later, a normalized expectation $E$ of energy
was proposed\cite{Tsal3,Penni} on the basis of the so called escort probability
\begin{equation} \label{a2}
E=\frac{\sum_{i}^w p_i^qe_i}{\sum_{i}^w p_i^q}.
\end{equation}
which, introduced into the maximization of Tsallis entropy as a
constraint replacing Eq.(\ref{aa1}), allows to obtain
\begin{equation} \label{a1a1}
p_i=\frac{[1-(1-q)\beta (e_i-E)]^\frac{1}{1-q}}{Z}
\end{equation}
with
\begin{equation} \label{a1a2}
Z=\sum_{i}^w[1-(1-q)\beta (e_i-E)]^\frac{1}{1-q}.
\end{equation}
The distribution Eq.(\ref{a1a1}) is invariant under energy
translation because, according to reference \cite{Tsal3}, if we
add a constant to $e_i$, we have the same constant added to $E$.
Recently, it is argued\cite{Marti} that, for a correct and
successful maximization of Tsallis entropy leading to
Eq.(\ref{a1a1}), the expectation Eq.(\ref{a2}) as a constraint
should be replaced by the unnormalized expectation Eq.(\ref{aa1})
minus the expectation given by Eq.(\ref{a2}), i.e.
$U-\sum_ip^qE=\sum_ip^q(e_i-E)$.

In this paper, we will firstly discuss different conditions for the validity of
energy translation invariance of probability distributions. Then we present
some observations about the relevant problems or theoretical peculiarities
within invariant NSM. The probability invariance with a generalized energy
shift consistent with NSM framework is discussed.

\section{Energy translation invariance of Maxwell-Boltzmann distribution}

As well known, BGS distribution is invariant under energy translation. This
property conforms with the intuition that the particle distributions in a
container will not be disturbed when, for example, you raise the container to a
higher point or move it at a different speed. This invariance of BGS is
directly related to that of classical mechanics which leaves the interaction
potential completely arbitrary and so is invariant if we add a constant to the
total energy. In addition, the exponential form of BGS allows following
invariant property :
\begin{equation} \label{b1}
p(e_i)=\frac{1}{Z}e^{-\beta e_i}=\frac{e^{-\beta C}}{e^{-\beta C}}
\frac{e^{-\beta e_i}}{\sum_j e^{-\beta e_j}}= \frac{e^{-\beta
(e_i+C)}}{\sum_j e^{-\beta (e_j+C)}}=p(e_i+C)
\end{equation}
So $p_i$ will not be changed if we add a constant to the energy spectrum $e_i$.
The partition function will be changed as follows :
\begin{equation} \label{b2}
Z(e+C)=\sum_je^{-\beta (e_j+C)}=e^{-\beta C}\sum_j e^{-\beta e_j}
=e^{-\beta C}Z(e)
\end{equation}
which gives following internal energy $U$ :
\begin{equation} \label{b3}
U(e+C)=-\frac{\partial}{\partial\beta}lnZ(e+C)=
-\frac{\partial}{\partial\beta}lnZ(e)+C = U(e)+C.
\end{equation}

It is instructive to see the role of thermodynamic limit in this
invariance when we study the velocity distribution of ideal gas.
Suppose that $\vec{v}_i$ is the velocity of the $l^{th}$ particle
of mass $m$ with respect to the container or to the center of
gravity of the gas. The total kinetic energy of the gas is given
by
\begin{equation} \label{b4}
e=\sum_l\frac{1}{2}m\vec{v}_l^2.
\end{equation}
Now we are in another system of reference in which the container moves at a
velocity $\vec{v}_0$, we have
\begin{equation} \label{b5}
e'=\sum_l\frac{1}{2}m(\vec{v}_0+\vec{v}_l)^2
=\frac{1}{2}m(\sum_l\vec{v}_l^2+\sum_l\vec{v}_0^2
+2\sum_l\vec{v}_0\cdot\vec{v}_l).
\end{equation}
In the thermodynamic limit, the third term at the right hand side of
Eq.(\ref{b5}) is null because $\sum_{l=1}^N\vec{v}_l=0$ if $N\rightarrow
\infty$. We then have $e'
=\frac{1}{2}m\sum_l\vec{v}_l^2+\frac{1}{2}mN\vec{v}_0^2
=e+\frac{1}{2}M\vec{v}_0^2$ where $M$ is the total mass of the gas. Put this in
Eq.(\ref{b1}) in replacing the summation by an integration in $v_l$, we find
that the distribution law does not change with $v_0$.

Briefly, according to the above discussion, we have at least three conditions
on the invariance of the distribution.

\begin{enumerate}
\item
The mechanics basis must be non relativist, i.e. an arbitrary
constant can be added to the energy.

\item
Thermodynamic limit holds, i.e. $N\rightarrow \infty$.

\item
The distribution law must be exponential.

\end{enumerate}

It is well known that the third condition needs the second one and, in
addition, the harsh assumptions that the interaction or correlation in the
system is negligible or of short term so that the different parts of the system
under consideration are independent and additive. In view of the above
mentioned conditions, it is difficult to say that the energy translation
invariance of probability distribution is an universal theoretical property.
Indeed, if we consider for example the relativistic effect, we can no more add
constant to the energy $e$ due to its relation with the total mass $M$ given by
\begin{equation} \label{b6}
e=Mc^2
\end{equation}
where $c$ is the light speed. $e$ is not arbitrary because $M$ can not be
changed arbitrarily. If we suppose a system composed of many elements of mass
$m_i$ and energy $\epsilon_i$ ($i=1, 2, 3, ...$) in interaction with total
potential energy $V$, $M$ will be given by :
\begin{equation} \label{b7}
M=\sum_im_i+V/c^2
\end{equation}
or
\begin{equation} \label{b7a}
Mc^2=\sum_im_ic^2+V=\sum_i\epsilon_i+V
\end{equation}
with $\epsilon_i=m_ic^2$ for $i^{th}$ element. It is obvious that $M$,
$\epsilon_i$ and $V$ cannot be changed, or the variance of the theory would be
perturbed\cite{Broglie}.

On the other hand, the relativistic kinetic energy of particles
($mc^2/\sqrt{1-v_l^2/c^2}$) is not linear function of $v_l^2$. So what we did
in Eq.(\ref{b5}) leading to $e'=e+mv_0^2/2$ does not hold for relativistic gas.
This means that the relativistic velocity distribution will change with
$\vec{v}_0$ even with exponential distributions.

Now if we consider complex systems with long range interactions which are not
limited between the walls of the containers, the things are more complex than
with short range interactions because the entropy and energy can be nonadditive
so that the exponential distribution does not exist any more. So the invariant
distribution is only a theoretical property in special cases. It should not be
claimed for general cases. It is worth emphasizing that, up to now, all the
successful applications of the distribution Eq.(\ref{a1}) are never disturbed
by the fact that it is not invariant with energy translation. On the contrary,
the invariant distributions like Eq.(\ref{a1a1}) were shown to present some
serious theoretical difficulties, as discussed below.

\section{Energy translation invariance of nonextensive distribution}
Nevertheless, there is up to now no any explicit reason to completely exclude
invariant nonextensive distribution which is recently discussed in a general
way and claimed to be an universal property verified by any thermostatistics
with linear or normalized expectation, whatever the entropy form\cite{Sist99}.
We have shown in the previous section that this conclusion is not true. But a
question remains open : is it possible to obtain the invariance with a
distribution function like Eq.(\ref{a1a1})? In this section, we would like to
show some observations about Eq.(\ref{a1a1}) and discuss some of its
theoretical peculiarities.

If we want a distribution $p_i$ to have energy translation invariance, we have
to first of all define an invariant normalized expectation satisfying
$U(e+C)=U(e)+C$. This is possible with the expectation Eq.(\ref{a2}) (or
Eq.(\ref{aa1}) plus the anomalous normalization
$\sum_{i}^wp_i^q=1$\cite{Wang00,Wang01,Wang02}) under the assumption that
$p_i(e_i+C)=p_i(e_i)$. That is
\begin{eqnarray} \label{g1}
E(e+C) &=& \sum_i(e_i+C)p_i^q(e_i+C)/\sum_ip_i^q(e_i+C) \\
\nonumber &=&\sum_i(e_i+C)p_i^q/\sum_ip_i^q \\
\nonumber &=&E+C
\end{eqnarray}

Then the method of \cite{Marti} can be adapted to introduce $\sum_{i}^wp_i^qE$
as the invariance constraint into the following functional

\begin{equation} \label{g2}
A=-\frac{\sum_{i=1}^{w}p_i-\sum_{i=1}^{w}p_i^q}{1-q}
-\alpha\sum_{i}^wp_i- \beta\sum_{i}^w p_i^qe_i + \gamma\sum_{i}^w
p_i^qE
\end{equation}
Let $\frac{\partial A}{\partial p_i}=0$, we obtain the following
distribution :
\begin{equation} \label{g3}
p_i=\frac{[1-(1-q)(\beta e_i-\gamma E)]^\frac{1}{1-q}}{Z}
\end{equation}
For $p_i$ to be invariant under energy translation, considering Eq.(\ref{g1}),
we have to set $\beta=\gamma$, which leads to the invariant distribution
Eq.(\ref{a1a1}).

\section{What is wrong with the invariant distribution?}

From the invariant distribution Eq.(\ref{a1a1}), we easily show that
\begin{equation} \label{f2a}
Z^{1-q}\sum_ip_i=\sum_ip_i^q[1-(1-q)\beta (e_i-E)]
\end{equation}
Due to the invariant factor $(e_i-E)$, Eq.(\ref{f2a}) leads to
\begin{equation} \label{f3a}
\sum_ip_i^q=Z^{1-q}
\end{equation}
and
\begin{equation} \label{f4a}
Z=\sum_i[1-(1-q)\beta (e_i-E)]^{\frac{q}{1-q}}.
\end{equation}
or
\begin{equation} \label{f5a}
\sum_i[1-(1-q)\beta (e_i-E)]^{\frac{1}{1-q}}
=\sum_i[1-(1-q)\beta(e_i-E)]^{\frac{q}{1-q}}.
\end{equation}

Eqs.(\ref{f3a}) and Eq.(\ref{f5a}) are obviously basic relations of the
invariant theory and should hold for arbitrary value of $q$, $\beta$ and $e_i$.
We will see that most of the problems encountered below are intrinsically
related to these two equalities.

\begin{enumerate}
\item
Let us begin by addressing the problem of the calculation of nonextensive
term or correlation in energy (or any other quantities of interest) with
Eq.(\ref{a1a1}).

Suppose an isolated system $C$ composed of two subsystems $A$ and $B$ in
thermal equilibrium. It was shown\cite{Abe01,Wang02} that, with Tsallis
entropy, the equilibrium condition yields for even interacting subsystems the
following product probability law
\begin{eqnarray} \label{f1}
p_{ij}(C)=p_i(A)p_j(B)
\end{eqnarray}
and entropy pseudoadditivity
\begin{equation} \label{f3}
S(A+B)=S(A)+S(B)+\frac{1-q}{k}S(A)S(B).
\end{equation}
For NSM with $q$-exponential distribution, the product probability means that
$A$ and $B$ are correlated and should give the pseudoadditivity associated with
the quantity of interest. But from Eqs. (\ref{a1a1}) and (\ref{f1}), we
straightforwardly obtain :
\begin{eqnarray} \label{f2}
e_{ij}(A+B)-E(A+B) & = &
[e_i(A)-E(A)]+ [e_j(B)-E(B)] \\
\nonumber
& + & (q-1)\beta[e_i(A)-E(A)][e_j(B)-E(B)].
\end{eqnarray}
Without additional hypothesis, this equality does not lead to any explicit
relation between the total energy and the energies of the subsystems $A$ and
$B$, which is absolutely necessary for defining temperature, pressure and
chemical potential as the measures of equilibrium.

Recently, some authors\cite{Abe99,Mart01} proposed neglecting the nonextensive
term in energy and writing
\begin{eqnarray} \label{f4}
e_{ij}(A+B) & = & e_i(A) + e_j(B)
\end{eqnarray}
and
\begin{eqnarray} \label{f5}
E(A+B) & = & E(A) + E(B).
\end{eqnarray}
We have seen that this additive energy approximation allows to reconcile the invariant theory with the zeroth
law of thermodynamic\cite{Abe99,Mart01} and to establish a generalized thermodynamics. But it should not be
forgotten that this is only an approximate approach. In other words, the invariant theory does not have
vigorously defined temperature. This is one of its intrinsic flaws.

\item
Now we address a mathematical problem related to Eqs.(\ref{f3a}) and (\ref{f5a}). From Eq. (\ref{f3a}),
Tsallis entropy can be recast as
\begin{equation} \label{f6}
S=k\frac{Z^{1-q}-1}{1-q}.
\end{equation}
Then we calculate the following derivative :
\begin{eqnarray} \label{f7}
\frac{dS}{dE}&=&\frac{k}{Z^q}\frac{dZ}{dE}.
\end{eqnarray}
First we take the $Z$ given by Eq. (\ref{a1a2}), we obtain :
\begin{eqnarray} \label{f8}
\frac{dS}{dE} &=& \frac{k}{Z^q}\frac{d}{dE}
\sum_{i}^w[1-(1-q)\beta(e_i-E)]^\frac{1}{1-q}\\ \nonumber
&=&\frac{k\beta}{Z^q}\sum_{i}^w[1-(1-q)\beta(e_i-E)]^\frac{q}{1-q}\\
\nonumber
&=&k\beta Z^{1-q}.
\end{eqnarray}
But considering Eq.(\ref{f5a}), we can also take the $Z$ of Eq.
(\ref{f4a}), this time we obtain
\begin{eqnarray} \label{f9}
\frac{dS}{dE}=\frac{qk\beta}{Z^q}\sum_{i}^w[1-(1-q)\beta
(e_i-U)]^\frac{2q-1}{1-q}.
\end{eqnarray}
Comparing Eq.(\ref{f9}) to Eq.(\ref{f8}), we get
\begin{eqnarray} \label{f10}
Z=q\sum_{i}^w[1-(1-q)\beta(e_i-E)]^\frac{2q-1}{1-q}.
\end{eqnarray}
If we put Eq. (\ref{f10}) into Eq. (\ref{f7}) and continue in this
way for $n$ times, we will find
\begin{eqnarray} \label{f11}
Z&=&\sum_{i}^w[1-(1-q)\beta(e_i-E)]^\frac{q}{1-q} \\
\nonumber
&=& q\sum_{i}^w[1-(1-q)\beta(e_i-E)]^\frac{2q-1}{1-q}\\
\nonumber
&=& q(2q-1)\sum_{i}^w[1-(1-q)\beta(e_i-E)]^\frac{3q-2}{1-q}\\
\nonumber
&=& q(2q-1)(3q-2)\sum_i^w[1-(1-q)\beta(e_i-E)]^\frac{4q-3}{1-q}\\
\nonumber &=& q(2q-1)(3q-2)...(nq-n+1)
\sum_i^w[1-(1-q)\beta(e_i-E)]^\frac{(n+1)q-n}{1-q}
\end{eqnarray}
with $n=0, 1, 2 ... $. We create in this way a series of equalities which seem
not to hold. For example, if we take the second equality of Eq. (\ref{f11}) and
let $q\rightarrow 0$, the right-hand side will tend to zero and the left-hand
side to $\sum_i^w1=w$. The result is $w\rightarrow 0$. This same result can
also be obtained for $q\rightarrow \frac{1}{2}$ if we take the third equality
of Eq. (\ref{f11}) and for $q\rightarrow \frac{2}{3}$ with the forth equality
and so on. These singular points in $q$ value do not conform with the
hypothesis that Eq. (\ref{f5a}) is a basic relation of the theory. It seems to
us that these equalities are valid only when $q\rightarrow 1$ and $Z$ becomes
the BGS partition function.

Now let us suppose a continuous quantity \^{x} within $0<x<\infty$ replacing
energy in the invariant distribution. Then $Z$ may be given by the following
integrations
\begin{eqnarray} \label{f12}
Z=\int_0^\infty [1-(1-q)\beta(x-\bar{x})]^\frac{1}{1-q}dx.
\end{eqnarray}
or
\begin{eqnarray} \label{f13}
Z=\int_0^\infty [1-(1-q)\beta(x-\bar{x})]^\frac{q}{1-q}dx.
\end{eqnarray}
In this case, we should put $q>1$ for $Z$ to be calculated when $x$ is large.
The integration of Eq.(\ref{f12}) is always finite. But Eq.(\ref{f13}) needs
$q<2$ to be finite. If $q>2$, the $Z$ of Eq.(\ref{f12}) can be calculated while
that of Eq.(\ref{f13}) diverges. This paradox naturally disappears for
$q\rightarrow 1$.

\item
The third problem concerns the calculation of expectation with
Eq.(\ref{a2}) and the concomitant distribution Eq.(\ref{a1a1}). Usually, this
calculation allows to establish a relation between, e.g., internal energy $E$
and micro-state energies $e_i$ through an $E-Z$ relationship
($U=-\frac{\partial}{\partial\beta}lnZ$ within BGS). This is the crucial step
in the statistical interpretation of thermodynamics. From Eq.(\ref{a1a1}), it
is obvious that $E$ calculation is self-referential : $E=O[p_i(E)]$ where
$C[.]$ is certain mathematical operation. This calculus does not make sense for
$E$ because Eq.(\ref{a1a1}) is itself a relative distribution with respect to
the internal energy $E$. So $E$ may be arbitrary for a given spectrum $e_i$. To
see this, we introduce the distribution function Eq.(\ref{a1a1}) into the
expectation Eq.(\ref{a2}) to obtain :
\begin{eqnarray} \label{f14}
E&=&\frac{\sum_{i} p_i^qe_i}{Z^{1-q}} \\ \nonumber
&=&\frac{1}{Z}\sum_{i}e_i[1-(1-q)\beta (e_i-E)]^\frac{q}{1-q}
\\ \nonumber &=&-\frac{1}{Z} \{ \frac{\partial}{\partial\beta}
\sum_{i}[1-(1-q)\beta (e_i-E)]^\frac{1}{1-q}
-\sum_{i}E[1-(1-q)\beta (e_i-E)]^\frac{q}{1-q} \}.
\end{eqnarray}
Considering Eqs.(\ref{f4a}) and (\ref{f5a}), we get
\begin{eqnarray} \label{f15}
E&=&-\frac{1}{Z} \{\frac{\partial Z}{\partial\beta}-EZ \} \\
\nonumber &=& -\frac{1}{Z}\frac{\partial Z}{\partial\beta}+E
\end{eqnarray}
which leads to, instead of the expected $U-Z$ relation,
\begin{eqnarray} \label{f16}
\frac{\partial Z}{\partial\beta} = 0.
\end{eqnarray}
So no relation between macroscopic quantities and the correspondent microscopic
quantities can be found from the expectation definition Eq.(\ref{a2}) and
arbitrary expectation is possible for any given states. The thermodynamics
connection of this statistical mechanics becomes questionable even impossible.

\item
In addition, Eq.(\ref{f16}) gives rise to another problem similar to the
second one discussed above. Eq.(\ref{f16}) can be easily verified if we take
the standard $Z$ given by Eq.(\ref{a1a2}). But If we take the $Z$ given by
Eq.(\ref{f4a}), we get the following relation

\begin{eqnarray} \label{f17}
\sum_{i}^w(e_i-E)[1-(1-q)\beta (e_i-E)]^\frac{2q-1}{1-q}=0
\end{eqnarray}
or
\begin{eqnarray} \label{f18}
E=\frac{\sum_{i}^we_i[1-(1-q)\beta
(e_i-E)]^\frac{2q-1}{1-q}}{\sum_{i}^w[1-(1-q)\beta
(e_i-E)]^\frac{2q-1}{1-q}}=
\frac{\sum_{i}^we_ip_i^{2q-1}}{\sum_{i}^wp_i^{2q-1}}.
\end{eqnarray}
If we repeat the same reasoning with the $Z$ of Eq.(\ref{f4a}), we get
\begin{eqnarray} \label{f19}
E& = &\frac{\sum_{i}^we_ip_i^q}{\sum_{i}^wp_i^q}
\\ \nonumber
&=&\frac{\sum_{i}^we_ip_i^{2q-1}}{\sum_{i}^wp_i^{2q-1}} \\
\nonumber &=&\frac{\sum_{i}^we_ip_i^{3q-2}}{\sum_{i}^wp_i^{3q-2}} \\
\nonumber &=&\frac{\sum_{i}^we_ip_i^{4q-3}}{\sum_{i}^wp_i^{4q-3}}
\\ \nonumber
...&=& \frac{\sum_{i}^we_ip_i^{nq-n+1}}{\sum_{i}^wp_i^{nq-n+1}}.
\end{eqnarray}
which means $\sum_{i}^w(e_i-E)=0$ or $E=\sum_{i}^we_i/w$ if $q=\frac{n-1}{n}$
with $n=1,2,3 ...$. i.e., we are led to the microcanonical case. This seems
contradictory to the generality of Eq.(\ref{f5a}).

\end{enumerate}

Above mathematical difficulties or peculiarities may lead to other difficulties
for the theory and seriously harm the reliability of NSM with invariant
distribution Eq.(\ref{a1a1}). The validity of escort probability and the
resulted expectation in NSM also become questionable because it gives
invariant distributions with respect to the expectation when applied to either
Tsallis' or R\'enyi's entropy \cite{Marti,Bash00}.

\section{An invariance recovered with a generalized energy shift}
Though Eq.(\ref{a1}) is not invariant with conventional energy shift ($e_i+C$),
it can be invariant with a different energy shift we refer to as generalized
energy translation.

We remember the product probability problem mentioned at the beginning of the
previous section. The existence of thermodynamic equilibrium in an interacting
system described by Tsallis entropy needs not only Eq.(\ref{f3}) and the
factorization of total probability given by Eq.(\ref{f1}), also an energy
pseudoadditivity similar to Eq.(\ref{f3}) and consistent with the product law
of probability Eq.(\ref{f1})\cite{Abe01,Wang02}, that is
\begin{equation} \label{h1}
e_{ij}(A+B)=e_i(A)+e_j(B)-(1-q)\beta e_i(A)e_j(B).
\end{equation}
and $p_{ij}[e_{ij}(A+B)]=p_i[e_i(A)]p_j[e_j(B)]$. Eq.(\ref{h1}) can be written
in the conventional addition form with a generalized $q$-addition ``$+_q$",
i.e. $e_{ij}(A+B)=e_i(A)+_qe_j(B)$. Let $q$-exponential distribution
Eq.(\ref{a1}) be noted by $p_i=exp_q(-\beta e_i)$, we get
\begin{eqnarray} \label{h2}
p_i(e_i+_qC) &=& \frac{1}{Z(e+C)}exp_q[-\beta (e_i+_qC)]
\\ \nonumber &=& \frac{1}{Z(e)Z(C)}exp_q(-\beta C)exp_q(-\beta e_i)=
\frac{1}{Z(e)}exp_q(-\beta e_i) \\ \nonumber &=& p_i(e_i).
\end{eqnarray}
where $Z(C)=exp_q(-\beta C)$ and $Z(e)=\sum_iexp_q(-\beta e_i)$. So the
invariance of probability is recovered in a more general mathematical and
physical context for systems with complex interactions or chaotic space time. A
complete physical comprehension of this ``nonlinear" energy shift certainly
needs further investigation.

Here we would like to mention that, in this generalized context, we can
construct a generalized arithmetic or algebra based on the $q$-exponential and
its inverse function $q$-logarithm ($\ln_q(x)=\frac{x^{1-q}-1}{1-q}$) with, in
addition to $+_q$, $-_q$, $\times_q$ and $\div_q$ (or $/_q$) corresponding
respectively to the following arithmetic properties : $exp_q(-_qx)=1/exp_q(x)$,
$\ln_q(x\times_qy)=\ln_q(x)+\ln_q(y)$ and $\ln_q(1/_qy)=-\ln_q(y)$. This
$q$-algebra is semi-classical due to its relationships containing
simultaneously classical and $q$-operations, but it is useful for NSM and
consistent with the complex physical circumstance. Considering the fact that
the classical arithmetic is constructed on the basis of {\it a fragmented world
with independent parts}, we conjecture that generalizations are necessary for
{\it the real messy world containing correlated, entangled and overlapped
parts}. We have already seen the development of a $\kappa$-algebra based on a
$\kappa$-statistics\cite{Kania01}. The above $q$-algebra is also a possible one
for a different physical circumstance. Detailed discussion of this mathematical
scenario will be given in another paper of ours.

\section{Conclusion}
We have discussed some problems related to the energy translation invariance of
probability distribution. It is shown that a probability distribution can be
invariant only in the thermodynamic limit and only if there is no long term
interaction or correlation and no relativistic effect is considered. So we
believe that this invariance is not to be imposed to nonextensive statistics
for complex systems containing long term interaction. Some theoretical
peculiarities of the invariant distribution reveal that the thermodynamic
connection of the NSM might be disturbed by this invariance. So the normalized
expectation Eq.(\ref{a2}) becomes problematic for NSM. In addition, the
distributions given by this expectation is numerically proved different from
that predicted by the law of large numbers\cite{Cour00}. All these problems
show that it is necessary to reconsider the role played by the escort
probability in NSM. It was shown that the invariant property of probability
distribution can exist with a generalized energy shift which is consistent with
the factorization of joint probability (product law of probability) prescribed
by the existence of thermodynamic equilibrium in correlated systems.

\acknowledgments We acknowledge with great pleasure the very useful discussions
with professor J.P. Badiali and Dr. F. Tsobnang.

\end{document}